# A Close Nuclear Black Hole Pair in the Spiral Galaxy NGC 3393

G. Fabbiano[1], Junfeng Wang[1], M. Elvis[1], G. Risaliti[1,2]

[1]Harvard-Smithsonian Center for Astrophysics (CfA). [2]INAF-Arcetri

The current picture of galaxy evolution[1] advocates co-evolution of galaxies and their nuclear massive black holes (MBHs), through accretion and merging. Quasar pairs (6,000-300,000 light-years[2,3] separation) exemplify the first stages of this gravitational interaction. The final stages, through binary MBHs and final collapse with gravitational wave emission, are consistent with the sub-light-year separation MBHs inferred from optical spectra[4] and light-variability[5] of two quasars. The double active nuclei of few nearby galaxies with disrupted morphology and intense star formation (e.g., NGC 6240[6] and Mkn 463[7]; ~2,400 and ~12,000 light-years separation respectively) demonstrate the importance of major mergers of equal mass spirals in this evolution, leading to an elliptical galaxy[8], as in the case of the double radio nucleus (~15 light-years separation) elliptical 0402+379[9]. Minor mergers of galaxies with a smaller companion should be a more common occurrence, evolving into spiral galaxies with active MBH pairs[10], but have hitherto not been seen. Here we report the presence of two active MBHs, separated by ~430 light-years, in the Seyfert[11] galaxy NGC 3393. The regular spiral morphology and predominantly old circum-nuclear stellar population[12] of this galaxy, and the closeness of the MBHs embedded in the bulge, suggest the result of minor merger evolution[10].

NGC 3393 (Table 1) was observed with *Chandra* ACIS-S on 2004-02-28 (ObsID 4868 for 29.7 ks) and 2011-03-12 (ObsID 12290 for 70 ks), giving 89.7 ks after screening for > 3σ background events. We used sub-pixel imaging with 1/4 the native 0.492''ACIS-S pixel, to recover the mirror resolution (~0.4" half maximum radius). We improved the *Chandra* and *HST* astrometry using stars in the field. Details are in the Supplementary Information. We used XSPEC for the spectral analysis, CIAO and DS9 for other analysis.

The image (fig. 1) shows a marginally extended source in the 3-8 keV band, suggesting some complexity of the emission. The spectrum shows a featureless 3-6 keV continuum and a prominent 6.4 keV Fe-K emission line, as previously reported[13]. We find that the continuum and line images differ: there is a single point-like source (NE) in the 3-6 keV band, coincident within errors with the *HST* nucleus, while the 6-7 keV image contains two sources with centroids 0.6" (~150 pc) apart. The ratio of the 6-7 keV and 3-6 kev images shows relatively more prominent Fe-K emission in the SW source, with position consistent with the VLBI maser[14]. The spectra of both sources (fig. 2) are typical of Compton-thick AGNs[13]. Fits to power-law and reflection component models (Table 2) show that both have observed (2-10 keV) luminosity of a few $10^{40}$ erg s$^{-1}$. The emitted luminosities estimated from the Fe-K line intensity[13] are $3.4\times10^{42}$ erg s$^{-1}$ (NE) and $5.0\times10^{42}$ erg s$^{-1}$ (SW). These high luminosities and spectral shapes exclude a starburst contribution, consistent with the predominantly old central stellar population[12].

If the 2-10 keV emission were solely from reflected nuclear emission in a Compton-thick AGN, the large dimensions of the reflector would preclude variability. However, variability is implied by the lower flux found in the *XMM-Newton* observation[15]. Transitions from Compton-thick to Compton-thin have been observed in some AGNs[16], suggesting temporary 'holes' in the wall of obscuring clouds. The *XMM-Newton* dimming may be related to the NE source, since an intrinsically obscured power-law component with Γ ~1.9 and $N_H$ ~ 2 x $10^{23}$ cm$^{-2}$ could fit its spectrum, suggesting that some direct nuclear emission may be visible. A passing broad line region cloud of $N_H$ ~ 2 x $10^{23}$ cm$^{-2}$ may have obscured this component during the *XMM-Newton* observation, leaving only residual scattered continuum. Even so, given the large Fe-K Equivalent Width (EW; Table 2), we may be only seeing reflected emission. In this case, we would be observing an additional absorption of $N_H$~2×$10^{23}$ cm$^{-2}$ towards the (warm) reflector.

The SW nuclear source, with weaker, flat continuum, and no optical Hα counterpart[17] (fig. 1), has prominent, and possibly complex, Fe-K emission. Although its luminosity is consistent with the most luminous ultra-luminous sources (ULXs) detected in nearby galaxies, its spectrum argues for a Compton-thick AGN. A 6.5 keV Fe line has been reported in the M82 X-1 ULX, but this line is very broad and its EW model-dependent[18]. The SW source Fe-K line, instead, has the well-defined narrow core found in Compton-thick nuclei, understood as fluorescent emission excited by the scattered nuclear radiation[19]. Moreover, unlike the SW source, strong continuum emission dominates the *Chandra* spectra of ULXs[20].

We can rule out that the two sources are the result of a single AGN interacting with clouds. Although the spectral shape of the emission of the SW source cannot exclude a local mirror reflecting flux from the NE source, given the ~150 pc separation, and the *Chandra* limit on size <0.4", the SW source covering factor is ≤1/70, implying an intrinsic luminosity of the NE source way above its measured value. Also, in the reflection hypothesis, we should be detecting Hα and [OIII] from the SW source, but none is seen[17]; this lack argues for a totally cocooned source, where the Fe-K emission is seen in transmission[13]. The spatial coincidence with the VLBI maser emission[14] (fig. 1) reinforces this conclusion. The NE source cannot be due to reflection by a similarly small reflector, given both its variability[15] and the modest far-IR luminosity of the system ~3 x $10^{43}$ erg s$^{-1}$ at 60 μm, based on the IRAS point source catalog flux of 2.25Jy, a value comparable with the estimated intrinsic nuclear emission. Finally, a jet interacting with the interstellar medium in the host galaxy bulge is also ruled out because the luminosity measured in the Fe-K line would require extreme conditions in terms of shock energy and gas ionization, and produce copious soft X-ray emission not seen with *Chandra.*

We conclude that there are two obscured AGNs in the central regions of NGC 3393. Both emitted luminosities inferred from the Fe-K lines are a few $10^{42}$ erg s$^{-1}$, showing that both sources contribute to the *BeppoSAX*[21] Compton-thick emission, with the SW source being more prominent.

The inferred intrinsic X-ray luminosities, for a standard AGN SED[22], and a quasar standard 10% accretion rate to luminosity conversion efficiency for the Eddington accretion rate, yield masses of $\sim 8 \times 10^5$ M$_\odot$ and $\sim 10^6$ M$_\odot$ for the NE and SW sources respectively. However, lower efficiencies and sub-Eddington accretion are possible, so the above masses are lower limits. The dynamical VLBI mass measurement[14] of $\sim 3 \times 10^7$ M$_\odot$ for the SW source implies that the product of efficiency and accretion rate must be lower by a factor of $\sim 30$.

If the masses of the two MBHs are similar, the MBH-bulge mass relation[23], would suggest that NGC 3393 may be the remnant of a major merger of two similar spiral galaxies. This merger would produce after $\sim 5$ Gyr a remnant with prominent grand design arms and SMB separation as seen in NGC 3393[8]. However, the stellar population of this bulge would be significantly rejuvenated (L. Mayer 2011, private communication), at odds with the age of the stellar population in the central 200 pc of NGC 3393[12]. Moreover, in a major merger the merging time scale for two MBHs at 150pc separation is $\sim$1Myr[8], making the detection of such events rare. Although this occurrence cannot be excluded on the base of a single detection, a better explanation of our results may be the merger of unequal mass galaxies (and MBHs), which would result in longer $\sim$1Gyr time scales[10], and would be consistent with the lack of widespread star formation. The constraints on the masses of the two MBHs allow this possibility. Interestingly, minor mergers may also result in the growth of one of the MBHs by promoting more active nuclear accretion in the smaller MBHs[17]. The denser circum-nuclear environment of the SW source, consistent with our results, may have resulted from such a process.

**Supplementary Information** is linked to the online version of the paper at www.nature.com/nature

**Acknowledgements.** This work was supported by a NASA Grant (PI: Junfeng Wang). We thank P. Gandhi for discussions on the 13μm observations.
We have used the NASA ADS and NED services. This works includes archival *Chandra* and *HST* data. We acknowledge useful discussions with Tiziana Di Matteo and Lucio Mayer at the Aspen Center for Physics 2011 Summer Program.

**Author Contributions.** GF suggested the possibility of double AGN, designed the data analysis approach, directed the interpretation of the results and wrote the paper. JW is the PI of the *Chandra* proposal, performed the data analysis and participated in the interpretation of the results. ME and GR contributed to the interpretation of the results and draft revisions.

**Author Information.** Reprints and permissions information is available at www.nature.com/reprints. Correspondence and requests for materials should be addressed to gfabbiano@cfa.harvard.edu

**Display Items Legends**

**Table 1. Summary of published results from X-ray observations of NGC 3393.** NGC 3393 (Distance ~50 Mpc; $L_H$~3.4×10$^{10}$L$_\odot$) has a Seyfert 2 nucleus[11], detected in H$\alpha$[17], 13micron[24] IR, and radio[14, 25]. The prominent 6.4 keV Fe-K$\alpha$ line and luminous ($L_X$~2 x 10$^{42}$ erg s$^{-1}$) high-energy X-ray (> 10 keV) emission[13, 15, 21, 26, 27, 28], suggest a Compton-thick active galactic nucleus (AGN). $L_X$ are observed values including continuum and Fe-K line emission, except when inferred from the Fe K or reflection component (PEXRAV) luminosities and so noted[13].

**Table 2. Results of the spectral analysis of Sources NE and SW.** The counts of each source were extracted in the 3-8 KeV band. Flux and luminosity were estimated in the 2-10 keV band to compare with results in the literature. Flux and luminosity are observed values including continuum and Fe-K line emission, except when inferred from the Fe K or reflection component (PEXRAV) luminosities and so noted, in which case they are estimates of the emitted values. Errors are 1$\sigma$ (68%). Aperture corrections were applied to account for the ~35% of the spectral photons in the wings of the *Chandra* PSF missed from the extraction regions. The total observed (2-10 keV) luminosity of the combined NE+SW emission is consistent with the results from the first *Chandra* observation[13], which did not resolve the two sources (see Table 1 for these comparisons). Considering the difference in beam size, and the luminosity of the rest of the galaxy[13], these fluxes are also consistent with the *Suzaku* measurement. The *XMM-Newton* measurement instead gives a lower value of the combined luminosity, suggesting a factor of 2 variability in the 2-10 keV range for the total NE+SW flux in 1year timescale. At higher energies (>15 keV), the 5 year Swift/BAT light-curve of NGC 3393[28] does not show any significant variability.

**Figure 1**. *Chandra* **ACIS-S images.** a) 3-8 keV image of the NGC 3393 nuclear region with ¼ subpixel binning, smoothed with a FWHM=0.25" Gaussian. Contours of *HST* F664N Hα[17] and *VLA* 8.4 GHz[25] emission are shown in gray and green, respectively. The diamond and cross indicate the position of the *HST*[17] and VLBI[14] sources, respectively. The X-ray source contains 279±16 counts in the 3-8 keV band. b) Image in the 3-6 keV spectral band, showing continuum emission dominated from NE source at RA= 10:48:23.47, Dec=-25:09:43.1 (all positions are J2000.0), coincident with the HST position[17]. This position is also consistent with that of the 13 μm source detected in the *VLT*/VISIR image[24] (also P. Gandhi 2011, private communication). c) Image in the 6-7 keV band including both continuum and Fe-K line emission; the dashed circles are the spectral counts extraction areas. d) Ratio of images shown in panels b) and c); the values of the ratios in the regions indicated by the dashed outlines are 0.61±0.04 (NW), 1.14±0.10 (SE) and 0.46±0.02 (in-between). The Fe-K emission is relatively more prominent in a source at the SW of the continuum source, at RA=10:48:23.45, Dec=-25:09:43.6. The latter position is closer to the nuclear position from the *Chandra* ObsID 4868[13], which was based on the centroid of the Fe-K emission. Given the astrometric uncertainty, the SW source is consistent with the VLBI position of the nuclear maser[14]. The sources are visible in both *Chandra* observations, although statistics are limited in the first. In the previous *Chandra* analysis[13], sub-pixel binning and imaging in separate spectral bands were not pursued.

**Figure 2** – **X-ray spectra.** (a) NE source, with intrinsically absorbed power-law best fit; (b) SW source, with best-fit PEXRAV model. Errors are 1 standard deviation. Spectra were extracted from the regions shown in fig. 1c with background from a source free region 10" to the east of the nuclei, and fitted them with XSPEC (version 12.6.0) using the C-Statistic. The NE spectrum, which shows a downturn around ~3 keV is fitted well with an intrinsically absorbed (in addition to line of sight Galactic $N_H \sim 4.7 \times 10^{20}$ cm$^{-2}$) power-law continuum plus Fe-K line, although the uncertainties in the parameters are large and this model is not a unique choice based on statistics. Given the flat continuum and more prominent Fe-K line, the SW source is likely to be Compton thick. An intrinsically absorbed power-law model fit gives only flat power laws and no intrinsic absorption. We have assumed a power-law + reflection component model for the continuum, fixing the $N_H$ to the Galactic value, that would be representative of Compton thick emission[13]. This model was also used for the NE source, as an alternative to the intrinsically absorbed power-law.

In the SW source the luminosity is dominated by the Fe-K emission. $L_X$(3-6 keV) = $3.7 \pm 0.7 \times 10^{39}$ erg s$^{-1}$, whereas $L_X$(Fe-K) = $1.2 \pm 0.3 \times 10^{40}$ erg s$^{-1}$. The equivalent width of the Fe-K line is 1.16 (+0.82, -0.58) keV (NE), and 2.77(+1.95, -1.43) keV (SW) at the 68% confidence level. The Fe-K line of the SW source appears broadened (0.13±0.05 keV) and complex, with a main peak at Fe-K$\alpha$ and a secondary peak at Fe-K$\beta$. The K$\beta$/K$\alpha$ ratio = 0.35±0.19, consistent with the expected range of I(K$\beta$)/I(K$\alpha$) ~ 0.12-0.2 for neutral or weakly ionized iron[29]. The results are listed in Table 2.

**Tables**

**Table 1 - Summary of published results from X-ray observations of NGC 3393**

| Observatory Date | Gamma $N_H$ (cm$^{-2}$) | X-ray band (keV) | Observed $F_X$ ($10^{-13}$ erg cm$^{-2}$ s$^{-1}$) | $L_X$ ($10^{40}$ erg s$^{-1}$) | I(FeK) ($10^{-6}$ ph cm$^{-2}$ s$^{-1}$) | EW(FeK) (keV) |
|---|---|---|---|---|---|---|
| *BeppoSAX*[21] Aug. 1, 1997 | 1.7 (fixed) >1×10$^{25}$ | 20-100 | 54 | 180 | 9.6 | 1.9(+3.8,-1.2) |
| *BeppoSAX*[15] Aug. 1, 1997 | 2.8(+1.2,-0.7) 4.4(+2.5, -1.1)×10$^{24}$ | | ... | ... | 14 | 4±2 |
| *XMM-Newton*[15] Jul. 5, 2003 | 1.6(+1.2,-1.2) >9×10$^{23}$ | 2-10 | 0.9(+0.6,-0.4) | 3.1(+2.1,-1.5) | 2.5 | 1.4±0.8 |
| *Chandra*[13] Feb. 28, 2004 | 1.9 (fixed) 4.7×10$^{20}$ (Gal.) | 2-10 | 2.1±0.04 | 7.5±0.1    720 (Fe K)    >100 (PEXRAV) | 4.2 | 1.4±0.7 |
| *Suzaku*[26, 27] May 23, 2007 | 1.52(+0.39,-0.38) 1.7(+1.4,-0.3)×10$^{24}$ | 2-10 <br> 15-50 | 4 <br> 200 | 14 <br> 680 | 4.3 | 0.5±0.2 |
| *Swift*[28] monitoring | 1.68(+0.30,-0.28) .... | 14-195 | 255 | 890 | | |

**Table 2. Results of the spectral analysis of Sources NE and SW**

| Source Cts. | Gamma $N_H$ (cm$^{-2}$) | Fx (2-10 keV) ($10^{-13}$ erg cm$^{-2}$ s$^{-1}$) | $L_X$ (2-10 keV) ($10^{40}$ erg s$^{-1}$) | I(FeK) ($10^{-6}$ ph cm$^{-2}$ s$^{-1}$) | $L_X$(FeK) ($10^{39}$ erg s$^{-1}$) | EW(FeK) (keV) | C-stat d.o.f |
|---|---|---|---|---|---|---|---|
| NE 134±9 | 1.9(+0.6,-1.8) (2.1±0.6)×10$^{23}$ | 1.3(+0.3,-0.1) | 4.4(+0.8,-0.5)    340 (Fe K) | 2.0±0.5 | 7±1.7 | 1.2(+0.8,-0.6) | 263 337 |
| | 1.7 (fixed) 4.7×10$^{20}$(Gal.) | 1.2±0.2 | 3.8±0.6    > 60 (PEXRAV) | 1.9±0.5 | 7±1.7 | 1.0(+0.7,-0.5) | 263 338 |
| SW 75±8 | 1.7 (fixed) 4.7×10$^{20}$ | 0.7(+0.2,-0.1) | 2.3(+0.7,-0.3)    500 (Fe K)    > 70 (PEXRAV) | 2.6±0.6 | 12.0±2.7 | 2.8(+1.9,-1.4) | 200 338 |

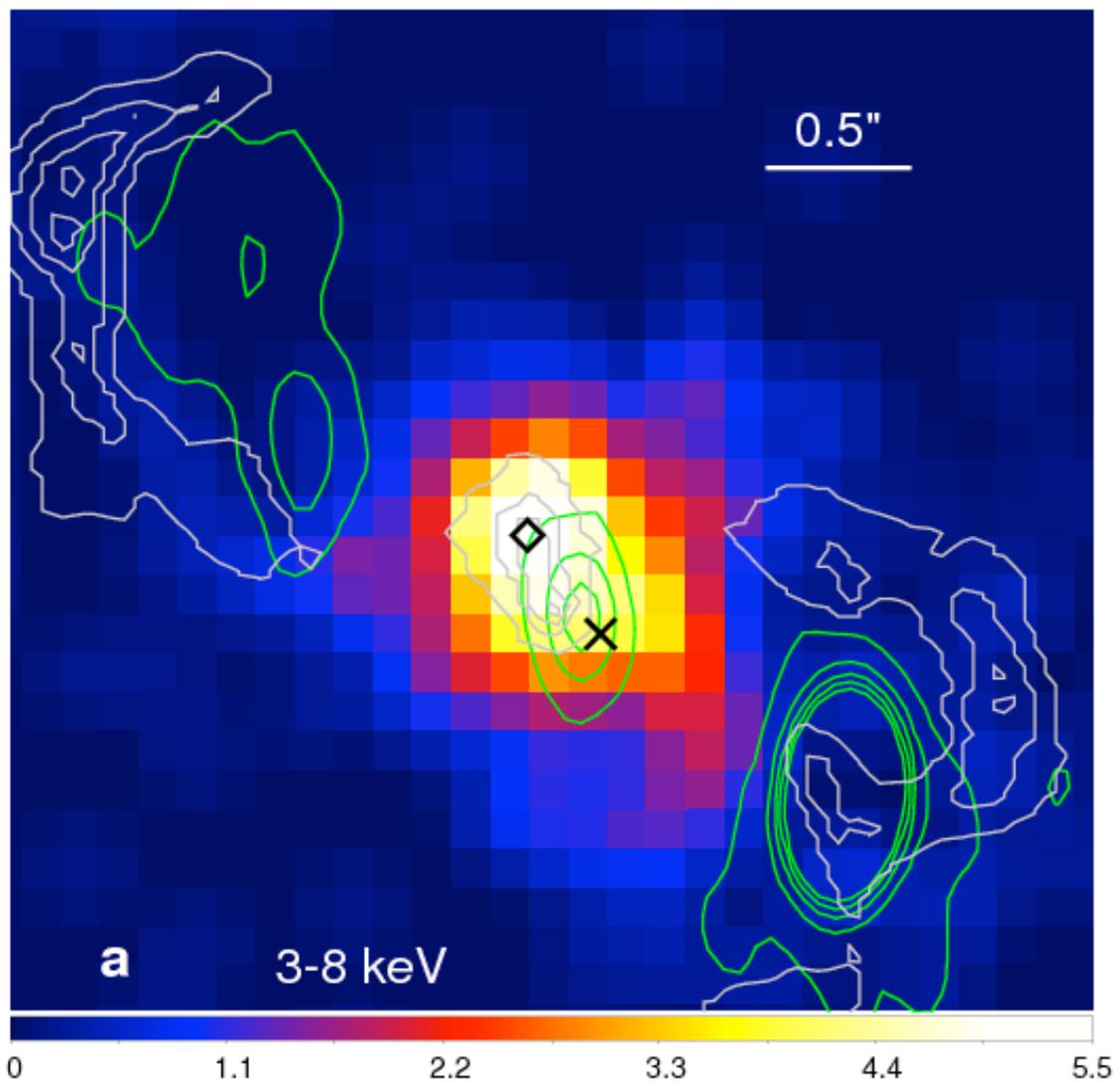

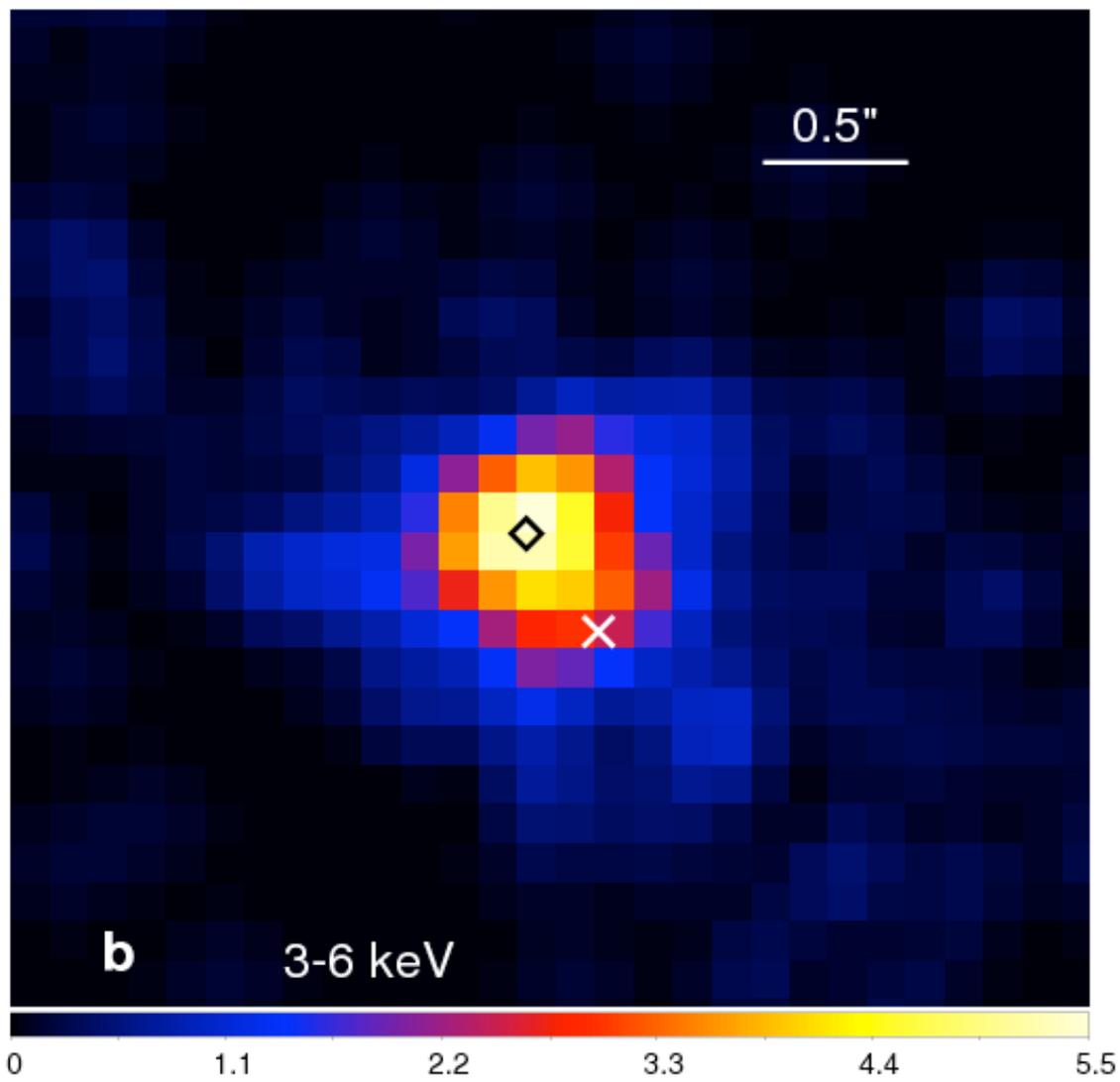

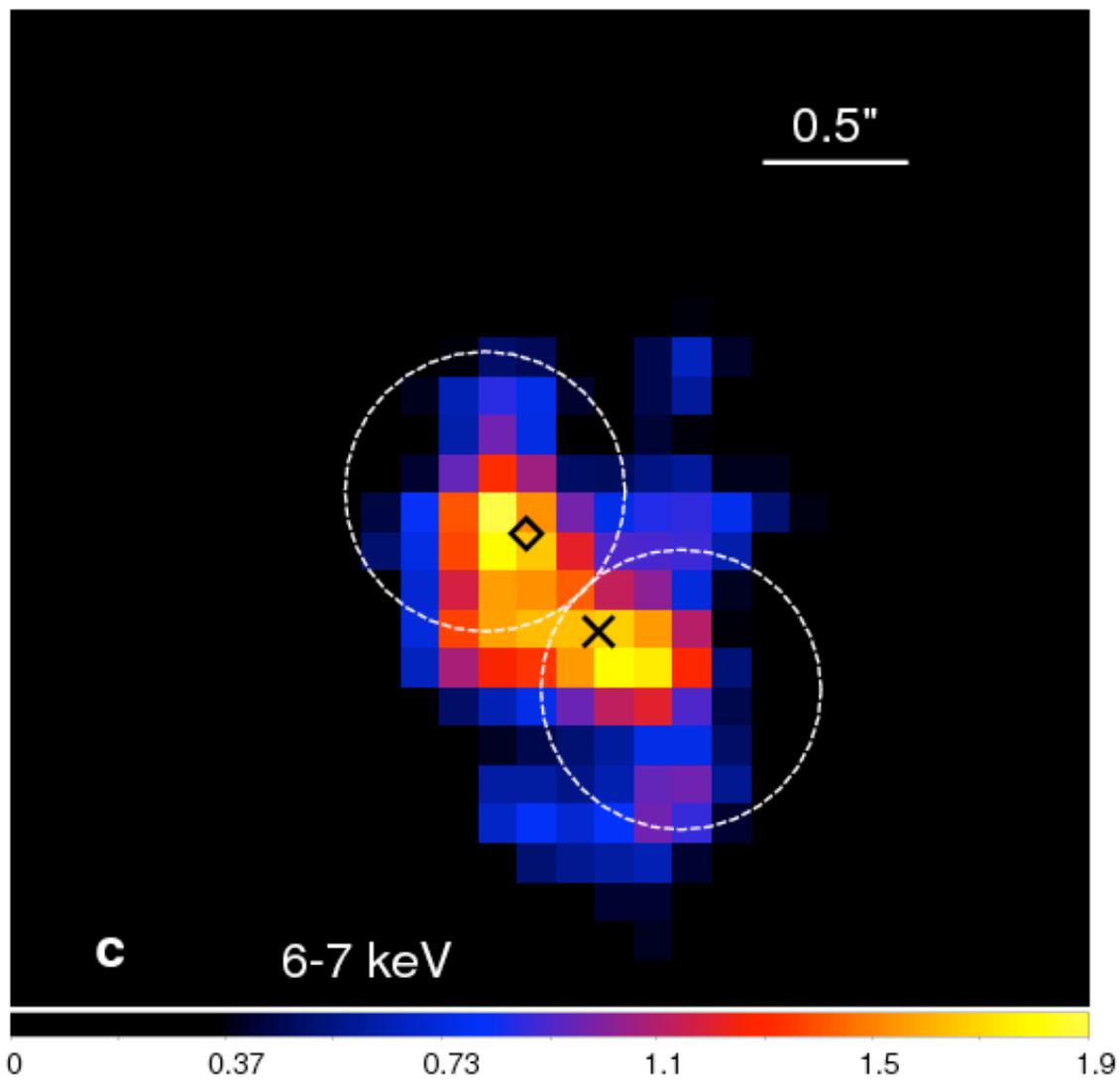

c  6-7 keV

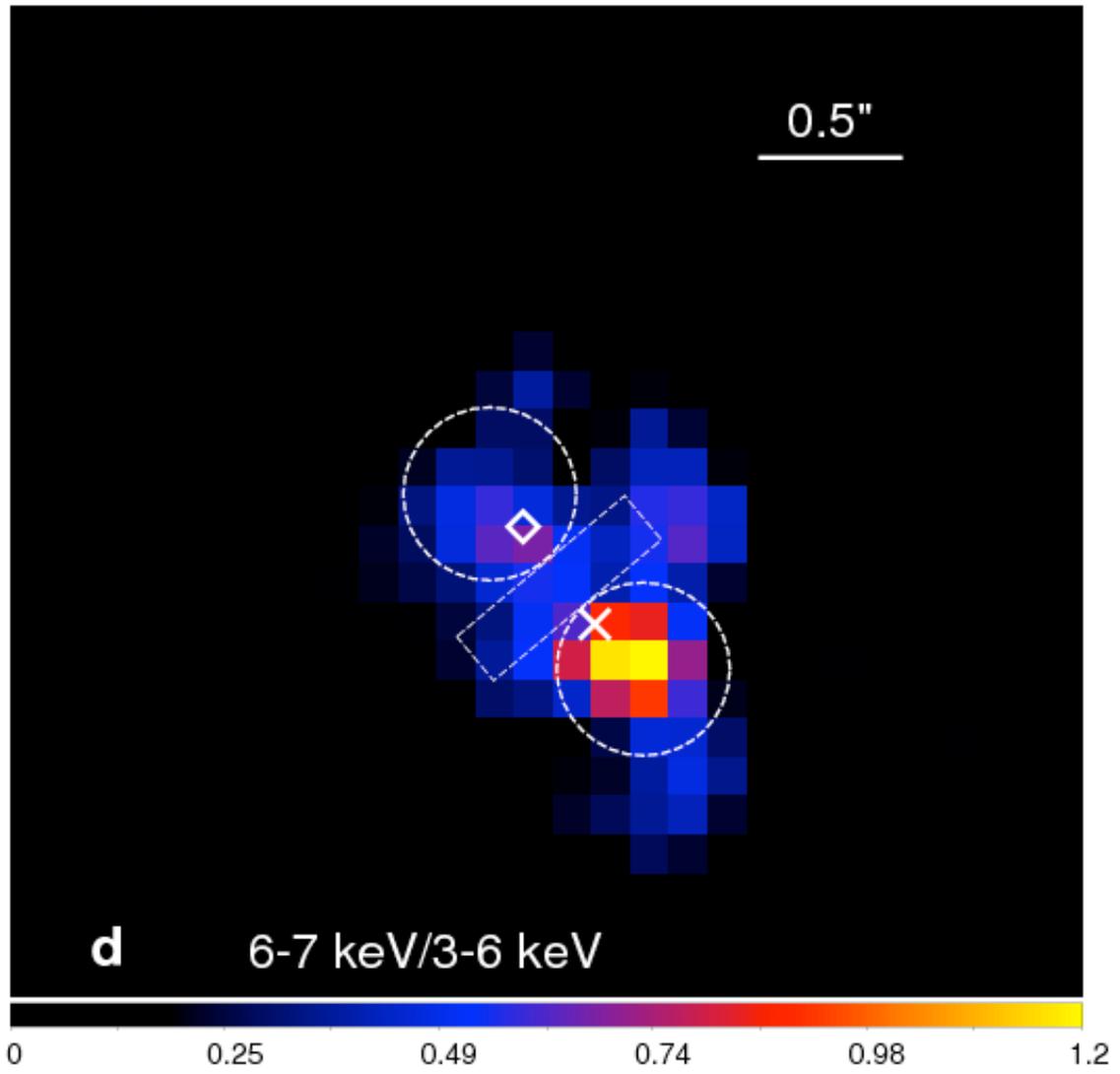

**Figure 1 a-d**

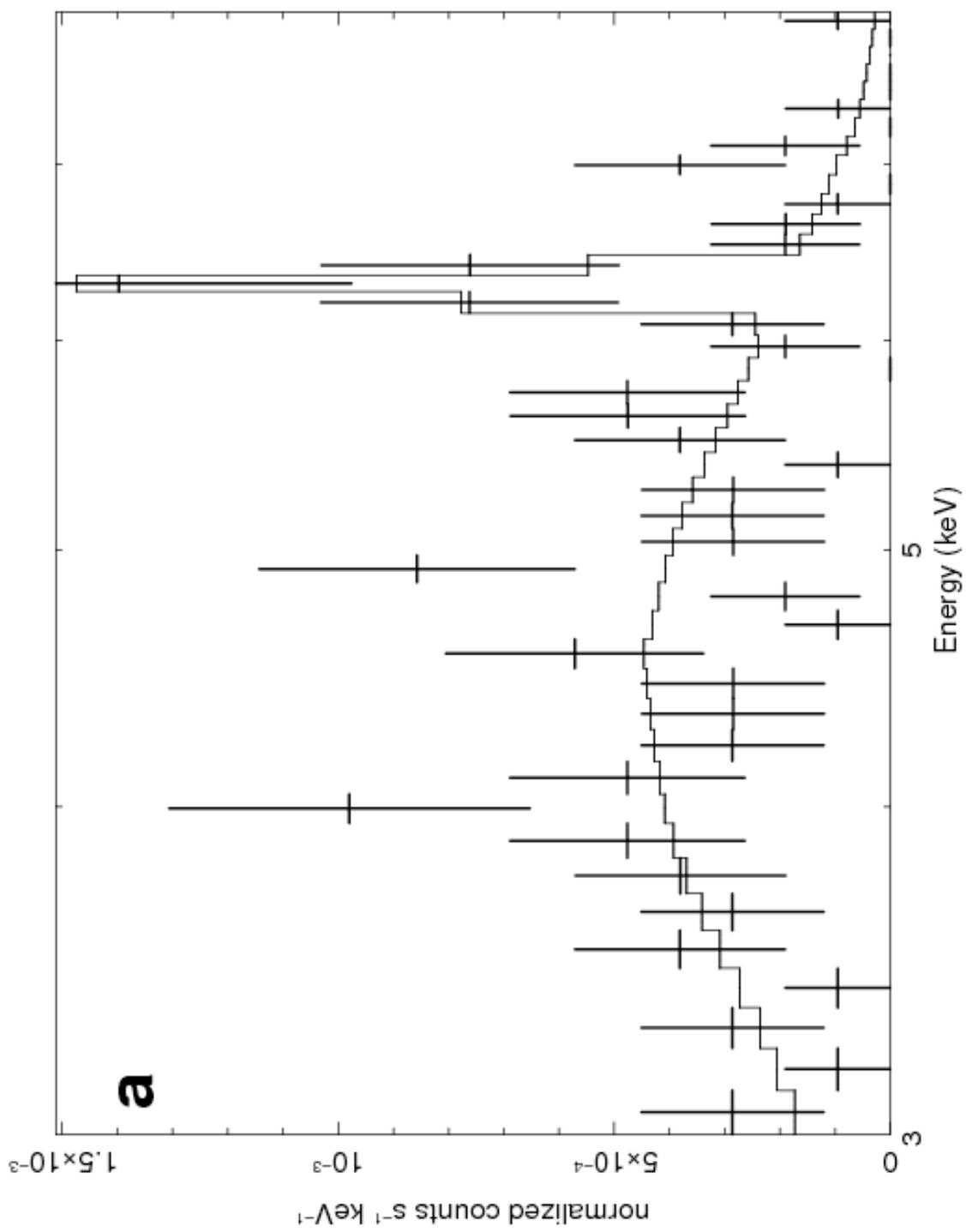

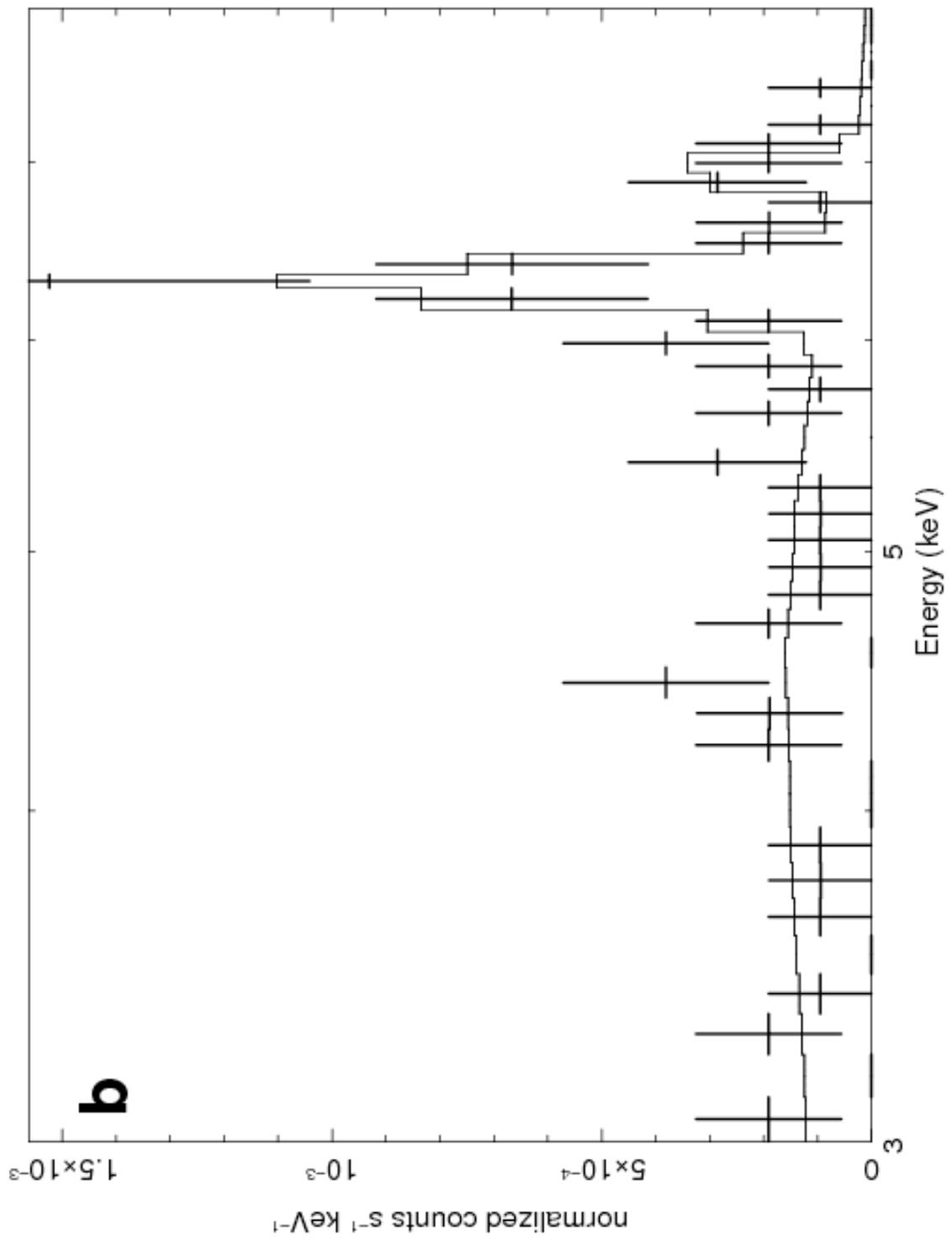

**Figure 2a-b**